\documentclass[oneside,a4paper]{article}
\usepackage[latin1]{inputenc}
\usepackage{enumerate,url}
\usepackage{amsfonts,amsmath,amssymb}
\usepackage{graphicx}
\usepackage{subcaption}
\usepackage{a4wide}

\usepackage{bm}
\usepackage[standard,thmmarks,thref]{ntheorem}
\usepackage[usenames,dvipsnames,svgnames,table]{xcolor}
\usepackage[normalem]{ulem}

\usepackage{arydshln} 
\usepackage{algorithm}
\usepackage{algorithmic}
\usepackage[obeyFinal]{todonotes}
\usepackage{setspace}

\usepackage{chicago}
\usepackage[colorlinks=true,citecolor=blue]{hyperref}

\newcommand{\tuple}[1]{\langle{#1}\rangle}
\definecolor{mygray}{RGB}{150,150,150}
\newcommand{\PCE}{\textsc{PCE-Coherence}}
\newcommand{\PCEx}{\textsc{PCE-Extension}}
\newcommand{\GNP}{\ensuremath{\mathrm{GNP}^s_k}}
\newcommand{\GNPk}{\ensuremath{\mathrm{GNP}_k}}
\newcommand{\dGNP}{\ensuremath{\mathrm{2GNP}}}

\renewcommand{\phi}{\varphi}
\renewcommand{\emptyset}{\varnothing}
\renewtheorem{example}{Example}[section]
\newtheorem*{M_Example}{Motivational Example}
\renewtheorem{lemma}[example]{Lemma}
\renewtheorem{definition}[example]{Definition}
\renewtheorem{proposition}[example]{Proposition}
\renewtheorem{theorem}[example]{Theorem}
\renewtheorem*{remark}[example]{Remark}

\title{Coherence of probabilistic constraints on Nash equilibria}

\author{Sandro Preto\\
	Institute of Mathematics and Statistics, University of São Paulo, Brazil\\
	spreto@ime.usp.br
	\and Eduardo Fermé\\
	University of Madeira and NOVA-LINCS, Portugal\\
	ferme@uma.pt
	\and Marcelo Finger\\
	Institute of Mathematics and Statistics, University of São Paulo, Brazil\\
	mfinger@ime.usp.br}

\date{}

\begin{document}
\doublespacing
	
\maketitle
	
\begin{abstract}
	\noindent Observable games are game situations that reach one of possibly many Nash equilibria.  Before an instance of the game starts, an external observer does not know, a priori, what is the exact profile of actions that will occur; thus, he assigns subjective probabilities to players' actions.  However, not all probabilistic assignments are coherent with a given game. We study the decision problem of determining if a given set of probabilistic constraints assigned a priori by the observer to a given game is coherent, which we call the Coherence of Probabilistic Constraints on Equilibria, or PCE-Coherence. We show several results concerning algorithms and complexity for PCE-Coherence when only pure Nash equilibria are considered. In this context, we also study the computation of maximal and minimal probabilistic constraints on actions that preserves coherence. Finally, we study these problems when mixed Nash equilibria are allowed.
	
\end{abstract}

\section{Introduction}
\label{sec:intro}

In game theory, a Nash equilibrium represents a situation in which each player's strategy is a best response to other players' strategies; thus no player can obtain gains by changing alone his/her own strategy.  Nash proved that every $n$-player, finite, non-cooperative game has a mixed-strategy equilibrium point~\cite{nash50a,nash50b,nash51}; however, more than one equilibrium may exist and the number of equilibria can be even exponentially large over some game parameters.

For an observer knowing that an equilibrium is to be reached, there is an a priori uncertainty before an instance of the game starts, concerning the exact kind of equilibrium to be reached and also in knowing the players' actions in that instance. Due to the computational complexity in explicitly producing the set of equilibrium states, the observer considers that set   as hidden or latent. Therefore, it is most natural to describe the observer's uncertainty in terms of subjective probabilities assigned to actions in those equilibrium states, in which one presupposes a probability distribution over the set of all possible equilibria. Our aim in this work is to study such a scenario comprehending a game and an observer, which we call an observable game.

We assume that the players involved in a game always react to each other with a best response in such a way that, since the game starts, its dynamics converge to an equilibrium. The intermediate steps where an equilibrium has not been achieved yet are not covered by the model, since the observer is only interested in the final stable outcome.
In case there are only two players and the first one to take an action does it smartly so that he will not need to change, the only source of randomness ends up being who will be the first player to take an action unless such action puts the second player in a position to choose among tied best responses. In any case, the proposed model aggregates all the possibilities of randomness concerning the actions to be taken. 

A way for the external observer to assign subjective probabilities to actions is looking at the past reactions of the players involved in a game situation that repeatedly happens. The configuration of the repeating games may even change from one instance to another regarding the allowed actions to each player or the utility values associated with the action profiles, however the observer may be able to grasp what kind of reaction is more likely to each player by observing the outcomes of the previous games.

Unfortunately, not every assignment on action probabilities by an observer finds correspondence to an actual probability distribution on possible equilibria of a given game; in fact, some actions may always co-occur at equilibrium, so constraining their probabilities to distinct values does not correspond to any underlying distribution on equilibria. In case the observer assigns a set of probabilistic constraints on actions that correspond to a probability distribution on equilibria, we say the observable game is coherent.

Lack of coherence can have important consequences which are better seen in a betting scenario where an observer knows the configuration of a game before one of its instances is played, and also knows that this game reaches an equilibrium. The observer wants to place bets on the occurrence of actions and an incoherent set of probabilities may lead to sure loss. So detecting and avoiding such a disastrous assignment of probabilities may have considerable cost to the observer. This betting scenario corresponds to de Finetti's interpretation of subjective probabilities \cite{dFi1931,dFi2017} in which incoherent probabilities have a one-to-one correspondence to sure loss. An actual scenario of this kind may be seen in the pricing strategy of oligopolistic markets.  We introduce the following motivational example.

\begin{M_Example}
	Assume that Madeira Beverages is a local beverages company in Madeira island that sells as its main product, beers. Two companies dominate the country beer market. They price their products from time to time in light of competition aiming to conquer the largest market share and make the most profit. Among the mechanisms of sale strategies there are price promotions (short-term price reductions), thus the price portfolio of a company in some period is not of public knowledge in advance. However, it is very reasonable to assume that the profits of the companies in the oligopoly reach an equilibrium during the sales period under consideration. Oblivious to the oligopoly competition, it is of great interest to Madeira Beverages to predict the price portfolios of the big companies based on its experience in observing their competition and pricing strategies; such prediction might help it to set up its own pricing strategy and even his production process, which takes place in a small and more limited industry. This information may be crucial, for example, for deciding to limit the production of beers that cannot be competitive with the oligopoly price portfolios of that period and focus on the production of some other beers with a more targeted niche or launch new non-beer products. In this scenario, the big companies, their price portfolios, and their profits (which may be inferred from their price portfolios), are respectively the players, their actions, and their utilities in a game; the local producer with predictions about the oligopolistic market is the observer with subjective probabilities over the player's actions.
\end{M_Example}

In this work, we formalize such scenario with a market in equilibrium and an external agent who has some idea about that equilibrium but is uncertain on the probabilistic  distribution on the possible equilibria and therefore on the players' actions. Of course, there may be aspects left out as it is expected from any theoretical idealization of the real world.

We study the following problems related to observable games according to pure and mixed Nash equilibrium over some different classes of games:
\begin{description}
	\item[The Coherence Problem] Given an observable game --- a game together with a set of probabilistic constraints on its actions ---, decide if it is coherent; that is, decide if there exists an actual probability distribution on the game equilibria that corresponds to those probabilistic constraints.
	\item[The Extension (Inference) Problem] Given a coherent observable game with probabilistic assignments on some of the players' actions, compute upper and lower bounds on the probabilities of some other action that preserves coherence.
\end{description}

We can divide the study of equilibrium problems in two fundamental aspects. On the one hand, the model \emph{per se}; in this case, equilibrium concepts are often understood as models that explain (rational or not) agent behavior, e.g. in the markets or in a biological system. On the other hand, algorithms; such issues become relevant in the cases where it is important to actually compute an equilibrium. Coherence of observable games --- or, coherence of probabilistic constraints on equilibria --- does not constitute a concept of equilibrium, however we can make an analogy between their study and the two aforementioned aspects. Indeed, the main goals of this work are:
\begin{itemize}
	\item Formalize the concept of coherent observable games; we also deepen the discussion on the kind of phenomena such concept model.
	\item Solve the Coherence and Extension Problems.
\end{itemize}
In the conclusions we propose an attempted explanation on how the algorithmic aspects may affect the phenomena modeled by (in)coherent observable games.

The combination of uncertainty and game equilibria is not trivial, so in order to better understand it we initially concentrate on uncertainty over pure equilibria, a restricted form of mixed-strategy equilibria in which each player chooses a unique action strategy, and whose existence is not even guaranteed. That is, the observer knows a priori that a pure equilibrium is to be reached for a given game, but does not know exactly which action strategies will be performed at equilibrium.

We later consider uncertainty in mixed-strategy equilibria, a doubly uncertain situation, that combines uncertainty on the actions to be played in a specific instance of a game with the probabilistic notion of mixed-strategy. It is important to note that the notion of probability on equilibria is different from  that of mixed-strategy equilibrium; in fact, these two notions are independent. The former deals with a probability distribution on possible equilibria, and the latter allows for probability distribution on an agent's actions as part of the player's strategy.

The paper is organized as follows. In Section \ref{sec:prelim} we define the preliminary notions of game, observable game, action profile, and pure Nash equilibrium. The concept of coherent observable game, and Coherence and Extension problems in the restricted setting of pure equilibria are formally defined in Section \ref{sec:model}. In Section \ref{sec:solving} we show the linear algebraic formulation of Coherence and Extension problems, we show some complexity and inapproximability results and propose solutions to the problems via reduction results; we also discuss the relations between Coherence and Probabilistic Satisfiability problems. In Section \ref{sec:pcemix} we generalize all the concepts and problems to the setting which includes mixed Nash equilibria; we discuss algorithms and show complexity and inapproximability results.

\section{Preliminaries}
\label{sec:prelim}

We define a \emph{game} as a quadruple $G = \tuple{P, N, A, u}$, where $P = \{1, \ldots, n\}$ lists the $n$ \emph{players} in the game, $N = \tuple{N_1, \ldots, N_n}$ is a sequence of player \emph{neighborhoods}, in which $N_i \subset P \setminus \{i\}$ is the set of player $i$ \emph{neighbors}, $A = A_1 \times \cdots \times A_n$ is a set of \emph{action profiles}, in which each $A_i$ is the set of all possible \emph{actions} for player $i$, and $u = \tuple{u_1, \ldots, u_n}$ is a sequence of \emph{utility functions} in which $u_i: A \to \mathbb{Q}$ is the utility function for player $i$. Assume that $A_i \cap A_j = \emptyset$ for player $i \neq j$, and that $u_i(a_1,\ldots,a'_j,\ldots,a_n)=u_i(a_1,\ldots,a''_j,\ldots,a_n)$ for $j \notin N_i\cup\{i\}$.

An action profile $e=\tuple{a_1, \ldots, a_i, \ldots, a_n}$ is a \emph{pure (Nash) equilibrium} if, for every player $i$, $u_i(e) \geq u_i(a_1, \ldots, a'_i, \ldots, a_n)$ for every $a'_i \in A_i$. A game $G$ may have zero or more pure equilibria.\footnote{Only mixed-strategy equilibria are guaranteed to exist, not pure ones; but every pure equilibrium is also a mixed-strategy equilibrium \cite{nash51}.} We write $a_i \in e$ to express that $a_i$ is the $i$th component of $e$.

By an \emph{observable game} we mean a pair $\mathcal{G} = \tuple{G,\Pi}$ where $G$ is a game and $\Pi$ is a set of \emph{probabilistic constraints on equilibria (PCE)}, that is a set of probability assignments on actions limiting the probabilities of some actions occurring in an equilibrium, which represents the observer's ignorance on what equilibrium will be reached; we assume it has the following format:
\[\Pi = \Big\{ P(\alpha_k) ~\bowtie_k~ p_k ~\Big|~ \bowtie_k \in \{\leq,\geq,=\}, 1 \leq k \leq K \Big\},\]
where $\alpha_k$ are actions and $p_k$ are values in $[0,1]\cap\mathbb{Q}$. Formally, a set of PCE is a set of triples $\Pi = \{\tuple{\alpha_k,\bowtie_k,p_k}, 1 \leq k \leq K\}$.

\section{Formalizing the concept}
\label{sec:model}

As observable games deal with the scenario where an equilibrium is to be reached but its action profile is unknown, we assign probabilities to equilibria:\footnote{N.B. Again, this probability function over equilibria should \emph{not} be confused with probabilities in mixed-strategies.} let $E_G = \{e_1, \ldots, e_M \}$ be the set of all equilibria associated with game $G$; we consider a probability function over $G$-equilibria $P: E_G \to [0,1] \cap \mathbb{Q}$, such that $P(e_i) \geq 0$ and $\sum_{e_i \in E_G} P(e_i) = 1$. We define the probability $P(a_i)$ that $a_i \in e \in E_G$ is executed in a game $G$ as
\[P(a_i) = \sum_{j \,|\, a_i \in e_j} P(e_j).\]

Given a game $G$ and an equilibrium probability function $P$, it is possible to compute the probability of any action; however we face two problems. First, the number of equilibria may be exponentially large in the numbers of players and of actions allowed for players. Second, we may not know the equilibrium probability function $P$. Instead we are presented with an observable game $\mathcal{G}=\tuple{G,\Pi}$, where $G$ is a game and $\Pi$ is a set of PCE and we are asked to decide the existence of an underlying probability function $P$ that satisfies $\Pi$; and, in case one exists, we want to compute the range of probabilities for an unconstrained action $a_i$. The former problem is called the \emph{probabilistic coherence problem} and the second one is the \emph{probabilistic extension problem}.

\begin{Definition}[PCE Coherence Problem]
	Given an observable game $\mathcal{G} = \tuple{G,\Pi}$, \PCE\ consists of deciding if it is \emph{coherent}, that is if there exists a probability function over the set of $G$-equilibria that satisfies all constraints in $\Pi$. \PCE\ rejects the instance if it is not coherent or if there exists no equilibrium in $G$.
\end{Definition}

\begin{Definition}[PCE Extension Problem]
	Let $\mathcal{G}$ be a coherent observable game. Given an action $a_i \in A_i$, \emph{\PCEx}\ consists in finding probability functions $\underline{P}$ and $\overline{P}$  that satisfy $\Pi$ such that $\underline{P}(a_i)$ is minimal and $\overline{P}(a_i)$ is maximal.
\end{Definition}

The coherence of an observable game models the interaction that the uncertainty about the game should have with the knowledge that such game reaches equilibrium. Thus, an incoherent observable game explains the inevitable failure of the observer in taking advantage of his position of observer, e.g. the sure loss of the better in de Finetti's probability interpretation or the poor management of the local beer producer observing the oligopolistic market. However, it is important to notice that the observer's subjective probability assignments may be coherent and still far from reality. In this way, an incoherent observable game alone may be enough to explain the failure of the observer, but a coherent observable game is not enough to guarantee his success. All in all, the sharpness of the observer's probability assignments also depends on how deep is his knowledge about the game and to be coherent is only part of his enterprise in making a good analysis of the game he observes.

\begin{Example} \label{ex:game}
	Suppose we have a game between Alice and Bob in which Alice has three possible actions $a^1$, $a^2$, and $a^3$, and Bob also has three possible actions $b^1$, $b^2$, and $b^3$, such that the joint utilities are given by Table \ref{tab:game}. This game has three pure Nash equilibria: $\tuple{a^1,b^1}$, $\tuple{a^2,b^3}$, and $\tuple{a^3,b^3}$, which are stressed in bold.
	\begin{table}[htb]
		\centering
		\begin{tabular}{c|ccc}
			& $b^1$ & $b^2$ & $b^3$ \\
			\hline
			$a^1$ & $\mathbf{2,2}$ & $1,1$ & $1,0$ \\
			$a^2$ & $1,2$ & $5,4$ & $\mathbf{1,5}$ \\
			$a^3$ & $0,1$ & $2,3$ & $\mathbf{1,3}$
		\end{tabular}
		\caption{Utility functions for Alice and Bob.}
		\label{tab:game}
	\end{table}
	Suppose the game will reach a pure equilibrium state, in which case Bob and Alice will have chosen to play a single action; we now want to see through an external observer's eyes who does not know which equilibrium will be reached, however gives to the action $a^2$ the probability of $\frac{1}{3}$. Is this restriction feasible (coherent)? And if it is, what is the lower bound on the probability of Bob playing $b^3$ this observer should assign in order to remain coherent? Can it be, say, $\frac{1}{4}$?
	
	Let us formalize such situation by $\mathcal{G}_1=\tuple{G_1,\Pi_1}$, where $G_1 = \tuple{P,N,A,u}$, in which $P = \{a,b\}$, $N_i = P\setminus\{i\}$, $A_a = \{a^1,a^2,a^3\}$, $A_b = \{b^1,b^2,b^3\}$, and $u$ is given by Table \ref{tab:game}. In $\Pi_1$, we consider the action $a^2$ occurring in an equilibrium with constraint $P(a^2) = \frac{1}{3}$. This constraint is coherent and it implies that the probability of $b^3$ is at least $\frac{1}{3}$. So if we consider $\Pi_1$ with joint constraints $P(a^2) = \frac{1}{3}$ and $P(b^3) = \frac{1}{4}$, $\mathcal{G}_1$ is incoherent.
\end{Example}

Observable game $\mathcal{G}_1$ in above example seems to imply that the formula $a^2 \to b^3$ holds, which forces $P(a^2 \to b^3)=1$ and thus $P(a^2) \leq P(b^3)$, so observable games can be seen as a particular encoding of a propositional logic theory.  Also note that in an $n$-player game, there may be exponentially many pure equilibria, posing a computational challenge to make explicit that encoded logic theory, and thus deciding coherence and finding upper and lower bounds for probabilities. Viewing observable games as implicit formulations of propositional theories motivates the following generalization of our goal problems.

Given a game $G$, consider a propositional logic language $\mathcal{L}_G$, whose atomic formulas are the actions $a_i \in A_i$; each such atomic formulas $a_i$ represents the occurrence of action $a_i$ in the equilibrium, and a formula $\varphi \in \mathcal{L}_G$ describes a Boolean combination of such statements at equilibrium. On the semantic side, each pure equilibrium $e$ defines a valuation $v_e$ such that for every action $a_i \in A_i$, $v_e(a_i) = 1$ iff $a_i \in e$.  So a formula $\varphi$ is satisfied at equilibrium $e$, represented as $\varphi \in e$, if $v_e(\varphi) =1$.

We can generalize the notion of PCE as a set of triples $\Pi = \{\tuple{\varphi_k,\bowtie_k,p_k}, 1 \leq k \leq K\}$, where $\varphi_k$ are formulas in this propositional language, so instead of restricting the probabilities of actions at equilibrium, we can now describe the probabilities of compound logical statements at equilibrium.

\begin{Definition}[Generalized PCE Problems]
	An observable game with a set $\Pi$ of generalized PCE is \emph{coherent} if there exists a probability function over the set of equilibria that satisfies all constraints in $\Pi$.  And the \emph{generalized PCE extension problem} for a coherent observable game with a generalized PCE $\Pi$ and a statement $\psi \in \mathcal{L}_G$ consists of finding upper and lower bounds for $P(\psi)$ that satisfy $\Pi$.
\end{Definition}

\section{Solving the problems}
\label{sec:solving}

This section concerns the algorithmic aspects of coherence of observable games. In this direction, in addition to the proposal of deterministic algorithms, we prove the following results for reasonable classes of observable games --- i.e. classes where the games have a compact representation.

\begin{itemize}
	\item If an observable game is coherent, then there is a probability distribution that assigns non-zero probability to a ``small'' number of equilibria that satisfies the observer's constraints. By ``small'' we mean $K+1$, where $K$ is the number of constraints assigned by the observer.
	
	\item The Coherence problem is in NP.
	
	\item If the decision on the existence of pure Nash equilibria is NP-complete, then the Coherence problem (considering only pure equilibria) is also an NP-complete problem.
	
	\item There cannot exist polynomial time (additive) approximation algorithms for solving the Extension problem with arbitrarily small precision $\varepsilon>0$, unless $P=NP$.
	
	\item For a given precision $\varepsilon > 0$, the solution of the Extension problem can be obtained with $O(|\log\varepsilon|)$ iterations of the Coherence problem.
\end{itemize}

\subsection{Game representations}
\label{sec:npgames}

We may find in the literature several ways to represent games, and this issue is directly related to the configuration of the instances for our problems and, thus, to its complexity classification. This work focuses on classes of games whose sizes are restricted and which possess equilibrium finding algorithms whose computation complexity is also restricted; we limit our attention to what we call \emph{GNP-classes}, in which the representation of the game takes polynomial space in the numbers $n$ of players and $s$ of maximum actions allowed for each player, and the computation of the pure equilibrium profiles may be made in non-deterministic polynomial time, also in terms of $n$ and $s$. Due to the time complexity restriction, the problem of deciding the existence of equilibria in a given GNP-class has complexity in NP.

We will always consider the problems \PCE\ and \PCEx\ with respect to a GNP-class. In other words, an instance of an observable game will be a pair $\mathcal{G}=\tuple{G,\Pi}$ where $G$ is a game in some GNP-class and $\Pi$ is a set of PCE in the form presented in Section \ref{sec:prelim} or generalized PCE as in previous section.

A natural way to represent games is by means of the \emph{standard normal form game} where the neighborhood of each player is $N_i = P \setminus \{i\}$, for all $i \in P$, and it is instantiated by explicitly giving its utility functions in a table with an entry for each action profile $a \in A$ containing a list with player utilities $u_i(a)$, for all $i \in P$, as in Example \ref{ex:game}.

It is an easy task to compute a pure Nash equilibrium of a game when its player utility functions are given extensively, as in standard normal form. In that case, we just need to check, for each action profile $e = \langle a_1, \ldots, a_i, \ldots, a_n \rangle$, whether it is a pure Nash equilibrium by comparing $u_i(e)$ with $u_i(a_1, \ldots, a'_i, \ldots, a_n)$, for all $i \in P$ and $a'_i \in A_i$. For each of the $|A|$ action profiles, $\sum_{i \in N} |A_i|$ comparisons will be needed. As the instance of the game is assumed to comprehend the utility function values for all players, the computation can be done in polynomial time in the size of the instance. However, in this explicit and complete format, the instance is exponential in the number $n$ of players, for if each player has exactly $s$ actions, each utility function has $s^n$ values and the game instance has $ns^n$ values to represent all utility functions.

Therefore, a class of standard normal form games fails to be a GNP-class since, despite equilibria being computable in polynomial time, the utility function requires exponential space to be explicitly represented.

More compact game representations, along with the complexity issues on deciding the existence of pure equilibria on them may be found, for example, in \cite{gottlob05}. We now introduce one of these compact representations in order to establish GNP-classes.

A \emph{graphical normal form game} is such that utility functions are extensively given in separate tables, for each player $i$, with an entry for each element in $\times_{j \in N_i \cup \{i\}} A_j$ containing a correspondent utility value $u_i(a)$, where it is enough to consider only the entries in $a$ with indices in $N_i \cup \{i\}$, since, as defined earlier, $u_i(a_1,\ldots,a'_j,\ldots,a_n)=u_i(a_1,\ldots,a''_j,\ldots,a_n)$ for $j \notin N_i\cup\{i\}$. Graphical normal form games can be turned into a compact representation by imposing the \emph{bounded neighborhood} property: let $k$ be a constant, we say that a game has $k$-bounded neighborhood if $|N_i| \leq k$, for all $i \in P$.

\begin{Example} \label{ex:gnf}
	Let $G_2 = \tuple{P,N,A,u}$ be a game with $P = \{a,b,c\}$, $A_a = \{a^1,a^2,a^3\}$, $A_b = \{b^1,b^2,b^3\}$, $A_c = \{c^1,c^2,c^3\}$, and utility functions given by Table \ref{tab:gnf}, from which on can infer the set $N$. $G_2$ is a game in graphical normal form with $k$-bounded neighborhood for $k\geq 1$, where for each player utility, only the previous player's action matters. As $k<n-1$, $G_2$ has a more compact representation than it would have in standard normal form. Note that this instance of graphical normal form game has $27$ utility values explicitly represented and the same game in standard normal form would need 81 utility values.
	\begin{table}
		\begin{subtable}{0.3\textwidth}
			\centering
			\begin{tabular}{c|ccc}
				& $a^1$ & $a^2$ & $a^3$ \\
				\hline
				$c^1$ & $10$ & $10$ & $5$ \\
				$c^2$ & $5$ & $10$ & $0$ \\
				$c^3$ & $5$ & $0$ & $10$
			\end{tabular}
		\end{subtable}
		~
		\begin{subtable}{0.3\textwidth}
			\centering
			\begin{tabular}{c|ccc}
				& $b^1$ & $b^2$ & $b^3$ \\
				\hline
				$a^1$ & $10$ & $5$ & $0$ \\
				$a^2$ & $10$ & $10$ & $5$ \\
				$a^3$ & $5$ & $0$ & $10$
			\end{tabular}
		\end{subtable}
		~
		\begin{subtable}{0.3\textwidth}
			\centering
			\begin{tabular}{c|ccc}
				& $c^1$ & $c^2$ & $c^3$ \\
				\hline
				$b^1$ & $10$ & $5$ & $0$ \\
				$b^2$ & $10$ & $10$ & $5$ \\
				$b^3$ & $5$ & $0$ & $10$
			\end{tabular}
		\end{subtable}
		\caption{Utility functions for players $a$, $b$, and $c$, respectively.}
		\label{tab:gnf}
	\end{table}
\end{Example}

It was shown that the problem of deciding whether a graphical normal form game has pure Nash equilibria is NP-complete \cite{gottlob05}, and that NP-hardness holds even when the game has $2$-bounded neighborhood, where each player can choose from only $2$ possible actions, and the utility functions range among $2$ values \cite{fischer06}. It is trivial to establish a non-deterministic polynomial algorithm for computing pure Nash equilibria on these games by guessing and then verifying it \cite{gottlob05,fischer06}.

Thus, we establish GNP-classes that contain the games in graphical normal form with $k$-bounded neighborhood and at most $s$ actions allowed to each player, for fixed $k \geq 2$ and $s \geq 2$; let \GNP\ represent these classes. Since it is needed at most $ns^k$ values to represent the utility functions, the representation of the games uses polynomial space in the number $n$ of players. Also, \GNPk $= \bigcup_{s\in\mathbb{N}}$ \GNP\ are GNP-classes where the representation of the games uses polynomial space in the number $n$ of players and the number $s$ of maximum actions allowed\footnote{It is also necessary that the length of utility function value representation be bounded by a polynomial in $n$ and $s$.}. Note that deciding the existence of pure equilibria in \GNP\ and \GNPk\ are NP-complete problems; we refer to the GNP-classes with this property as NP-complete GNP-classes.

\subsubsection{Computing Nash equilibria via Boolean satisfiability}

The Cook-Levin Theorem~\cite{Coo71} guarantees that there exists a polynomial reduction from the problem of computing pure equilibria on GNP-classes to Boolean Satisfiability (SAT). Next, we are going to show such a reduction.

Given a game $G$ with $P = \{1, \ldots, n\}$ and $A_i = \{a_i^1, \ldots, a_i^{s_i}\}$, for $i \in P$, we build a CNF Boolean formula $\phi_G$ with the variables $x_i^j$ meaning that player $i$ chose action $a_i^j$. Let $k$ be the maximal $|N_i|$ and $s$ be the maximal $|A_i|$, $s_i \leq s$. The formula $\phi_G$ is a set of clauses as follows:
\begin{enumerate}[(a)]
	\item For each player $i$, add a clause $\bigvee_{j=1,\ldots,s_i} x_i^j$, representing that each player chooses one action. This set of clauses is built in time $O(ns)$.
	\item For each player $i$ and pair $a_i^p$, $a_i^q$, with $p \neq q$, add a clause $\neg x_i^p \vee \neg x_i^q$, representing that each player chooses only one action. This set of clauses is built in time $O\left(n {s \choose 2}\right)$.
	\item \label{clauses_equilibrium} For each player $i$ and $a = \tuple{a^{q_1}_1, \ldots, a^{q_{i-1}}_{i-1}, a^{q_{i+1}}_{i+1}, \ldots, a^{q_n}_{n}}$, add the clause $\bigvee_{j \in N_i} \neg x_j^{q_j} \vee \bigvee_{r \in R} x_i^r$, where $R$ is the set of indices $r$ such that $u_i(a^{q_1}_1, \ldots, a^{q_{i-1}}_{i-1}, a_i^r, a^{q_{i+1}}_{i+1}, \ldots, a^{q_n}_{n}) \geq u_i(a^{q_1}_1, \ldots, a^{q_{i-1}}_{i-1}, a'_i, a^{q_{i+1}}_{i+1}, \ldots, a^{q_n}_{n})$, for all $a'_i \in A_i$, representing each player chooses one of the best responses depending on his neighborhood choices; there may be more than one best response all of which have the same utility. This set of clauses is built in time $O(ns^k)$.
\end{enumerate}
For games in \GNP, $\phi_G$ is built in linear time in $n$, and for games in \GNPk, it is built in polynomial time in $n$ and $s$. A non-deterministic polynomial algorithm for computing pure Nash equilibria consists of the aforementioned reduction from a game $G$ to its corresponding Boolean formula $\varphi_G$ followed by an NP algorithm computing a satisfiable valuation for $\varphi_G$; the valuations satisfying $\varphi_G$ naturally encode action profiles that are pure Nash equilibria.


\begin{Example} \label{ex:gnf_phi}
	For the game $G_2$ in Example \ref{ex:gnf} the CNF formula $\phi_{G_2}$ contains the variables $x_a^1$, $x_a^2$, $x_a^3$, $x_b^1$, $x_b^2$, $x_b^3$, $x_c^1$, $x_c^2$, $x_c^3$ and the clauses:
	\medskip
	\begin{enumerate}[(a)]
		\item $x_a^1 \vee x_a^2 \vee x_a^3$, $x_b^1 \vee x_b^2 \vee x_b^3$, $x_c^1 \vee x_c^2 \vee x_c^3$; \smallskip
		\item $\neg x_a^1 \vee \neg x_a^2$, $\neg x_a^1 \vee \neg x_a^3$, $\neg x_a^2 \vee \neg x_a^3$, $\neg x_b^1 \vee \neg x_b^2$, $\neg x_b^1 \vee \neg x_b^3$, $\neg x_b^2 \vee \neg x_b^3$, $\neg x_c^1 \vee \neg x_c^2$, $\neg x_c^1 \vee \neg x_c^3$, $\neg x_c^2 \vee \neg x_c^3$; \smallskip
		\item $\neg x_c^1 \vee x_a^1 \vee x_a^2$, $\neg x_c^2 \vee x_a^2$, $\neg x_c^3 \vee x_a^3$, $\neg x_a^1 \vee x_b^1$, $\neg x_a^2 \vee x_b^1 \vee x_b^2$, $\neg x_a^3 \vee x_b^3$, $\neg x_b^1 \vee x_c^1$, $\neg x_b^2 \vee x_c^1 \vee x_c^2$, $\neg x_b^3 \vee x_c^3$.
	\end{enumerate}
\end{Example}

\subsection{Complexity of \PCE}
\label{sec:pcepure}

We start by formulating \PCE\ in linear algebraic terms. Let $\mathcal{G}=\tuple{G,\Pi}$ be an observable game where $G$ is a game with $M$ pure Nash equilibria and $\Pi = \{ P(\alpha_i) \bowtie_i p_i, 1 \leq i \leq K \}$ is a set of PCE; consider a $K\times M$ matrix $A = [a_{ij}]$ such that $a_{ij}=1$ if $\alpha_i \in e$, where $e$ is the $j$-th pure Nash equilibrium of $G$, and $a_{ij}=0$ otherwise. Then, \PCE\ is to decide if there is a probability vector $\pi$ of dimension $M$ that obeys:
\begin{eqnarray}
\nonumber
A \pi &\bowtie& p\\
\label{eq:PCErestrictions}
\mbox{$\sum \pi_j$} &=& 1\\
\nonumber
\pi &\geq& 0
\end{eqnarray}

Note that, if we were dealing with a generalized PCE $\Pi = \{ P(\varphi_i) \bowtie_i p_i, 1 \leq i \leq K \}$, the linear algebraic formulation would be exactly the same, with the following remark.
An equilibrium $e$ satisfies a formula $\varphi$ ($\varphi \in e$) if the Boolean valuation $v_e$ that satisfies only the propositional atoms which are actions in $e$ ($v_e(\alpha_i)=1$ iff $\alpha_i \in e$) also satisfies $\varphi$ ($v_e(\varphi)=1$).  So, from this point on, we do not make any distinction between generalized PCE and its basic version.

The observable game $\mathcal{G}=\tuple{G,\Pi}$ is coherent if there is a vector $\pi$ that satisfies \eqref{eq:PCErestrictions}. We join the first two conditions in \eqref{eq:PCErestrictions} in just one matrix $A$. Of course, it is not mandatory for the \PCE\ instance to attach a constraint to each action, in which case matrix $A$ has fewer lines than the number of actions involved. The next results establish computational complexity for \PCE.

\begin{Theorem} \label{thm:nppurenp}
	\PCE\ over a GNP-class is a problem in NP.
\end{Theorem}

\begin{proof}
	Suppose the observable game $\mathcal{G}=\tuple{G,\Pi}$ is coherent and $|\Pi|=K$. Therefore there exists a probability distribution $\pi$ over the set of all possible pure Nash equilibria that satisfy $\Pi$. By the Carath\'eodory's Theorem~\cite{Eck93} there is a probability distribution assigning non-zero probabilities to at most $K+1$ equilibria. These equilibria are polynomially bounded in size since $G$ is member of a GNP-class. Therefore, there is a witness $\pi$ whose size is polynomially bounded attesting $\Pi$ is satisfied, so $\mathcal{G}$ is coherent and \PCE\ is in NP.
\end{proof}

\begin{Theorem} \label{thm:nppurehard}
	\PCE\ over an NP-complete GNP-class is NP-complete.
\end{Theorem}

\begin{proof}
	Membership in NP follows from Theorem \ref{thm:nppurenp}. For NP-hardness, let us reduce the problem of deciding the existence of pure Nash equilibria for games in the NP-complete GNP-class at hand to \PCE\ over this same class. Given a game $G=\tuple{P,N,A,u}$, we consider the instance of observable game $\mathcal{G}=\tuple{G,\{P(a_i)\geq 0\}}$, for some arbitrary $a_i\in A_i$, for $i\in P$. The reduction from $G$ to $\mathcal{G}$ may be computed in linear time; and $\mathcal{G}$ is coherent if, and only if, $G$ has a pure Nash equilibrium. We have shown that \PCE\ is NP-hard.
\end{proof}

\begin{Corollary}
	\PCE\ over \GNP\ and \GNPk\ are NP-complete.
\end{Corollary}

\subsection{An algorithm for \PCE}

We provide an algorithm for solving \PCE\ by means of a reduction from this problem to the probabilistic satisfiability problem (PSAT), which is a well studied problem for which there already are algorithms and implementations in the literature~\cite{GKP88,HJ2000,FDB2011,DF2015b}.

\subsubsection{Probabilistic satisfiability}
\label{sec:psat}

A \emph{PSAT instance} is a set $\Sigma=\{P(\alpha_i) \bowtie_i p_i ~|~ 1 \leq i \leq K\}$, where $\alpha_1, \ldots, \alpha_k$ are classical propositional formulas defined on $n$ logical variables $\mathcal{P} = \{x_1, \ldots, x_n\}$, which are restricted by probability assignments $P(\alpha_i) \bowtie_i p_i$, $\bowtie_i\, \in \{=, \leq, \geq\}$ and $1 \leq i \leq K$. Probabilistic satisfiability consists in determining if that set of constraints is consistent, defined as follows.

Consider the $2^n$ possible propositional valuations $v$ over the logical variables, $v: \mathcal{P} \rightarrow \{0,1\}$; each such valuation is extended, as usual, to all formulas, $v: \mathcal{L} \rightarrow \{0,1\}$. A \emph{probability function over propositional 	valuations} $\pi: V \rightarrow [0,1]$ is a function that maps every propositional valuation to a value in the real interval $[0,1]$ such that $\sum_{i=1}^{2^n} \pi(v_i) = 1$.  The probability of a formula $\alpha$ according to function $\pi$ is given by $P_\pi(\alpha) = \sum \{\pi(v_i) ~|~ v_i(\alpha) = 1\}$.

Nilsson~\cite{Nil86} provides a linear algebraic formulation of PSAT, consisting of a $K \times 2^n$ matrix $A = [a_{ij}]$ such that $a_{ij} = v_j(\alpha_i)$. The \emph{probabilistic satisfiability problem} is to decide if there is a probability vector $\pi$ of dimension $2^n$ that obeys the \emph{PSAT restriction}:
\begin{align}
\nonumber
A \pi &\bowtie p\\
\mbox{$\sum \pi_j$} &= 1 \label{eq:PSATrestrictions}\\
\nonumber
\pi &\geq 0
\end{align}

If there is a probability function $\pi$ that solves \eqref{eq:PSATrestrictions}, we say $\pi$ satisfies $\Sigma$. In such a setting, we define a PSAT instance $\Sigma$ as \emph{satisfiable} if \eqref{eq:PSATrestrictions} is such that there is a $\pi$ that satisfies it. Clearly, the conditions in \eqref{eq:PSATrestrictions} ensure $\pi$ is a probability function. Usually the first two conditions of \eqref{eq:PSATrestrictions} are joined, $A$ is a $(k+1) \times 2^n$ matrix with 1's at its first line, $p_1=1$ in vector $p_{(k+1) 	\times 1}$, so $\bowtie_1$-relation is ``$=$''.

As a consequence of Carath\'eodory's Theorem~\cite{Eck93}, it is provable that any satisfiable PSAT instance has a ``small'' witness.

\begin{Proposition}[\cite{GKP88}] \label{fact:NP}
	If a PSAT instance $\Sigma=\{P(\alpha_i) \bowtie_i p_i ~|~ 1 \leq i \leq k\}$ has a solution $\pi$ satisfying~\eqref{eq:PSATrestrictions}, then there is a solution $\pi'$ also satisfying~\eqref{eq:PSATrestrictions} such that $\pi'_j > 0$ for at most $k+1$ elements; the remaining elements of $\pi'$ are thus 0.
\end{Proposition}

The existence of a small witness in Proposition~\ref{fact:NP} serves as an NP-certificate for a satisfiable instance, so PSAT is in NP. Furthermore, note that by making all probabilities $1$ in \eqref{eq:PSATrestrictions}, the problem becomes a simple \emph{propositional satisfiability} (SAT), so PSAT is NP-hard. It follows that PSAT is NP-complete.

\subsubsection{Reducing \PCE\ to PSAT}

The reduction proceeds as follows: consider an observable game $\mathcal{G}=\tuple{G,\Pi}$ where $G$ is member of a GNP-class and $\Pi = \{ P(\alpha_i) = p_i, 1 \leq i \leq K \}$ is a set of PCE. We now construct a PSAT instance $\Sigma_\mathcal{G}$ as follows. As $G$ is in a GNP-class, we have shown in Section~\ref{sec:npgames} that there exists a classical propositional formula $\phi_G$ over a set of atoms $x_i, \ldots, x_N$, $N \leq ns$, representing all possible actions in $G$, such that if $\phi_G$ is satisfied by valuation $v$, $v(\phi_G)=1$, then $\{x_i ~|~ v(x_i)=1 \}$ is a set of atoms representing actions in equilibrium. Furthermore, the actions $\alpha_i$ that appear in $\Pi$ may also be considered as atoms in $x_1, \ldots, x_N$. Make $\Sigma_\mathcal{G} = \Pi \cup \{ P(\phi_G)=1 \}$.

\begin{Theorem} \label{thm:nppurepsat}
	Let $\mathcal{G}=\tuple{G,\Pi}$ be an observable game where $G$ is a member of a GNP-class and $\Pi$ is a set of PCE, and $\Sigma_\mathcal{G}$ be its associated PSAT instance constructed from $\mathcal{G}$ as above. Then, $\mathcal{G}$ is coherent if, and only if, $\Sigma_\mathcal{G}$ is satisfiable.
\end{Theorem}

\begin{proof}
	Suppose $\mathcal{G}$ coherent. There exists a probability distribution $P$ over the set of equilibria $E_G = \{e_1, \ldots, e_M\}$ such that $\sum_{j=1}^M P(e_j) = 1$ and that satisfies $\Pi$. Since each equilibrium is associated with a valuation that takes value $1$ in the atoms $x_i$ within the equilibrium and $0$ otherwise, we consider the probability distribution over valuations as the probability distribution over equilibria, taking probability $0$ to those valuations which are not associated to equilibria. This probability distribution makes $\Sigma_\mathcal{G}$ satisfiable.
	
	Now, suppose $\Sigma_\mathcal{G}$ satisfiable. As the probability distribution that satisfies $\Sigma_\mathcal{G}$ makes $P(\phi_G) = 1$, it has non-zero value only on valuations related to equilibria. Since it also satisfies $\Pi$, considering this distribution as a probability distribution over equilibria, we find $\mathcal{G}$ coherent.
\end{proof}

\begin{Corollary}
	\PCE\ over a GNP-class is polynomial time reducible to PSAT.
\end{Corollary}

\begin{Remark}
	Since PSAT is in NP, it follows from Theorem \ref{thm:nppurepsat} that \PCE\ is also in NP. In other words, Theorem \ref{thm:nppurenp} can be seen as a corollary of Theorem~\ref{thm:nppurepsat}.
\end{Remark}

\begin{Example} \label{ex:pcepure}
	We show the reduction of \PCE\ to PSAT for the observable game $\mathcal{G}_2=\tuple{G_2,\Pi_2}$ with $G_2$ in Example \ref{ex:gnf} and $\Pi_2$ a set of PCE consisting of vector $p$ below. We omit the columns of matrix $A$ in PSAT restriction~\eqref{eq:PSATrestrictions} that represent valuations which do not satisfy $\phi_{G_2}$, calculated in Example \ref{ex:gnf_phi}. So, the columns in matrix $A$ codify the five pure Nash equilibria in $G_2$; its first line represents $\sum \pi_i = 1$.
	\begin{align*}
	A\pi =
	\begin{array}{c}
	\phantom{c} \\ a^1 \\ a^2 \\ a^3 \\ b^1 \\ b^2 \\ b^3 \\ c^1 \\ c^2 \\ c^3 \\ \phi_{G_2}
	\end{array}
	\left[
	\begin{array}{ccccc}
	1 & 1 & 1 & 1 & 1\\
	1 & 0 & 0 & 0 & 0\\
	0 & 1 & 0 & 1 & 1\\
	0 & 0 & 1 & 0 & 0\\
	1 & 0 & 0 & 1 & 0\\
	0 & 1 & 0 & 0 & 1\\
	0 & 0 & 1 & 0 & 0\\
	1 & 0 & 0 & 1 & 1\\
	0 & 1 & 0 & 0 & 0\\
	0 & 0 & 1 & 0 & 0\\
	1 & 1 & 1 & 1 & 1
	\end{array}
	\right] \cdot \left[
	\begin{array}{c}
	\pi_1 \\ \pi_2 \\ \pi_3 \\ \pi_4 \\ \pi_5
	\end{array}
	\right] = \left[
	\begin{array}{c}
	1 \\ 0.1 \\ 0.9 \\ 0 \\ 0.5 \\ 0.5 \\ 0 \\ 0.8 \\ 0.2 \\ 0 \\ 1
	\end{array}
	\right] = p
	\end{align*}
	This PSAT instance is satisfiable due to, for example, the vector $\pi = [0.1, 0.2, 0, 0.4, 0.3]'$, so the \PCE\ instance is coherent.
\end{Example}

\subsubsection{Phase transition phenomenon}
\label{sec:phasettrans}

The reduction from \PCE\ to PSAT is particularly interesting due to the phenomenon of phase transition, which is observed in the implementation of PSAT-solving algorithms.

\begin{figure}[ht]
	\centering
	\includegraphics[width=.7\textwidth]{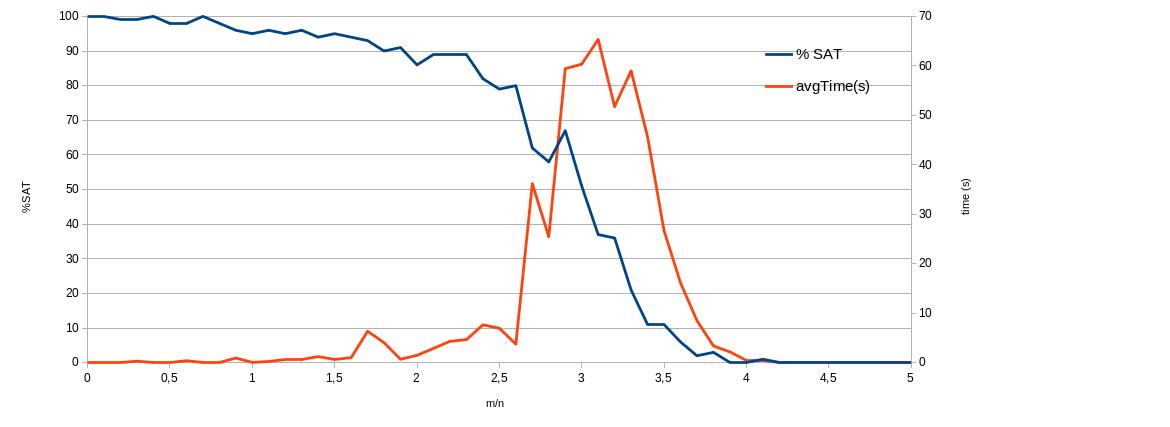}
	\caption{PSAT Phase Transition; each point was obtained by solving 100 instances of PSAT in the same class.}
	\label{fig:psatPhaseTrans}
\end{figure}

Phase transition is an empirically observable property of practical solutions of a decision problem. It starts by imposing a linear order on classes of instances of given problem; for example,  in 3-SAT,  one may fix the number $n$ of propositional variables, so each class consists of 3-SAT instances with $\frac{m}{n}$ clauses. A first-order phase transition occurs when one moves from mostly positive decisions in a class to mostly negative decisions;  in 3-SAT  this can be verified,  as one moves from  problems with very few constraints (clauses), so that almost any valuation satisfies an instance in the class,  to problems with too many constraints, mostly unsatisfiable ones, as illustrated by the blue curve in Figure~\ref{fig:psatPhaseTrans}. A second-order phase transition occurs when one observes the average time taken to solve an instance; it is reasonably small both for unconstrained or very constrained   instances,  but it sharply increases when one approaches from both  ends  the point where basically 50\% of  the instances are decided positively, which is called the phase transition point.  This second order phase-transition is empirically observable but has no  theoretical explanation  in the case of  NP-complete problems, as illustrated by the red curve in Figure~\ref{fig:psatPhaseTrans}. When the number $n$ of variables increases, it is observed that the phase transition point remains fixed, but the slopes become more sharp.

The existence of a second-order phase transition for all NP-complete problems was conjectured by~\cite{CKT1991},  and it was demonstrated to hold for the  SAT  problem and related problems~\cite{KS1992,GW1994}, which are problems that show an easy-hard-easy empirical complexity, with most instances in the easy parts.   In particular,  several  algorithms for the PSAT problem  have  displayed such behavior~\cite{FDB2011,DF2015b}, such as the one in \cite{DF2015b} presented in Figure~\ref{fig:psatPhaseTrans}.

Thus, the reduction from \PCE\ to PSAT in encouraging, specially when regarding classes \GNP, where such reduction is linear, because many of the resulting PSAT-instances are expected to be solved easily.

\subsection{An algorithm for \PCEx}

Let us turn to the \PCEx\ problem. Given a coherent observable game $\mathcal{G}$, our aim is to find the maximum and minimum observer's probabilistic constraints for some action $\alpha$ maintaining coherence. In other words, we need to search among the NP-witnesses of $\mathcal{G}$ for some that maximizes and minimizes the constraints on $\alpha$. One might wonder whether there are polynomial time (additive) approximation algorithms for such problem, i.e., given a \PCEx\ instance consisting of $\mathcal{G}$ and $\alpha$ and a precision $\varepsilon>0$, whether there exist polynomial time algorithms which return $m$ and $M$ such that:
\begin{itemize}
	\item $|\underline{P}(\alpha) - m | < \varepsilon$;
	\item $|\overline{P}(\alpha) - M | < \varepsilon$.
\end{itemize}
The next results show the answer is negative, unless a huge breakthrough in complexity theory is achieved. First we establish an auxiliary reduction: from a game $G=\tuple{P,N,A,u}$, we build the game $G^*=\tuple{P,N,A^*,u^*}$, where $A^*_1=A_1\cup\{b\}$, with $b\notin A_1$, and $A^*_i=A_i$, for $i\in P\setminus\{1\}$. Profiles $a\in A\subset A^*$ remain with the same utilities $u^*_i(a)=u_i(a)$, for all $i\in P$, and new profiles $p_b=\tuple{b,a_2,\ldots,a_n}\in A^*$, have utilities $u^*_1(p_b)=\max\{ u_1(a'_1,a_2,\ldots,a_n) ~|~ a'_1\in A_1 \}$ and $u^*_i(p_b) = \max\{ u_i(a) ~|~ a\in A \}$, for $i\in P\setminus\{1\}$.

\begin{Lemma}\label{lemma:puregamereduction}
	Game $G^*$ may be built from a game $G$ in polynomial time and has the new pure Nash equilibria $p_b$ in addition to the ones $G$ already has.
\end{Lemma}

\begin{proof}
	Game $G^*$ may be built in polynomial time because for every partitioning set $\{ \tuple{x,a_2,\ldots,a_n} ~|~ x\in A_1 \}$ of action profiles of $G$, we may add one unique new action profile $p_b=\tuple{b,a_2,\ldots,a_n}$; then it is necessary to add to $G^*$ less new utility values than the description of $G$ already has. Let $a=\tuple{a_1,\ldots,a_n}\in A$ be an action profile. If $a$ is a pure Nash equilibrium of $G$, players in $P\setminus\{1\}$ cannot increase their utilities by choosing other action in $A^*_i$ and, if player $1$ were able to do so, it would have to be by choosing action $b$, then $u^*_1(b,a_2,\ldots,a_n) > u_1(a'_1,a_2,\ldots,a_n)$, for all $a'_1\in A_1$, contradicting the definition of $u^*_1$. If $a$ is not a pure Nash equilibrium of $G$, all players can increase their utilities by choosing other actions in $A_i$. Then, all pure Nash equilibria in $G$ remains pure Nash equilibria in $G^*$. Finally, action profiles $p_b=\tuple{b,a_2,\ldots,a_n}$ are clearly pure Nash equilibria in $G^*$ and we have the result.
\end{proof}

\begin{Theorem}\label{thm:pureapproxmin}
	Unless $P=NP$, there does not exist a polynomial time algorithm that approximates, to any precision $\varepsilon\in(0,\frac{1}{2})$, the expected value by the minimization version of \PCEx.
\end{Theorem}

\begin{proof}
	Deciding the existence of pure Nash equilibria for games in \GNPk\ is an NP-complete problem; let us reduce this problem to \PCEx. Given a game $G$, we consider the coherent observable game $\mathcal{G}=\tuple{G^*,\{P(a_i)\geq 0\}}$, for some arbitrary $a_i\in A_i$, for $i\in P$, together with action $b$ as an instance of \PCEx. The reduction from $G$ to $\mathcal{G}$ may be computed in polynomial time by Lemma \ref{lemma:puregamereduction}. Suppose there exists a polynomial time algorithm that approximates to precision $\varepsilon\in(0,\frac{1}{2})$ the expected value by the minimization version of \PCEx. If $G$ does not have any pure Nash equilibrium, all equilibria in $G^*$ are of the type $p_b=\tuple{b,a_2,\ldots,a_n}$, then $\underline{P}(b)=1$ and the supposed algorithm should return $m>1-\varepsilon>\frac{1}{2}$. On the other hand, if $G$ has some pure Nash equilibrium, $\underline{P}(b)=0$ and the supposed algorithm should return $m<0+\varepsilon<\frac{1}{2}$. Therefore, such algorithm decides an NP-complete problem in polynomial time and $P=NP$.
\end{proof}

\begin{Theorem}\label{thm:pureapproxmax}
	Unless $P=NP$, there does not exist a polynomial time algorithm that approximates, to any precision $\varepsilon\in(0,\frac{1}{2})$, the expected value by the maximization version of \PCEx.
\end{Theorem}

\begin{proof}
	Deciding the existence of pure Nash equilibria for games in \GNPk\ is an NP-complete problem; let us reduce this problem to some instances of \PCEx. Given a game $G$, we consider the coherent observable game $\mathcal{G}=\tuple{G^*,\{P(a_i)\geq 0\}}$, for some arbitrary $a_i\in A_i$, for $i\in P\setminus\{1\}$, together with all actions $a_1\in A_1$ as $|A_1|$ different instances of \PCEx. The reduction from $G$ to $\mathcal{G}$ may be computed in polynomial time by Lemma \ref{lemma:puregamereduction}. Suppose there exists a polynomial time algorithm that approximates to precision $\varepsilon\in(0,\frac{1}{2})$ the expected value by the maximization version of \PCEx. If $G$ does not have any pure Nash equilibrium, all equilibria in $G^*$ are of type $p_b=\tuple{b,a_2,\ldots,a_n}$, then $\overline{P}(a_1)=0$, for all $a_1\in A_1$, and the supposed algorithm should return $M< 0+\varepsilon<\frac{1}{2}$, for all \PCEx\ instances concerning $a_1\in A_1$. On the other hand, if $G$ has some pure Nash equilibrium, $\overline{P}(a_1)=1$, for some $a_1\in A_1$, and the supposed algorithm should return $M>1-\varepsilon>\frac{1}{2}$, for a particular \PCEx\ instance concerning some $a_1\in A_1$. Therefore, we are able to decide the existence of a pure Nash equilibrium in game $G$ by running the supposed algorithm $|A_1|$ times in the instances comprehending $\mathcal{G}$ and $a_1\in A_1$; $G$ has a pure Nash equilibrium, if it returns $M>\frac{1}{2}$ for some instance, and $G$ has no equilibrium otherwise. Such routine based on the supposed algorithm decides an NP-complete problem in polynomial time, hence $P=NP$.
\end{proof}

In the following, by means of a simple application of the binary search algorithm, we are able to provide a deterministic procedure to solve \PCEx, which shows that its complexity burden is all due to \PCE. Given a precision $\varepsilon=2^{-k}$, we proceed by making a binary search through the binary representation of the possible constraints to $\alpha$, solving \PCE\ in each step. Algorithm \ref{alg:pcex} presents the procedure to solve the maximization version of \PCEx. We called $\mathit{PCECoherence(G,\Pi)}$ the process that decides a \PCE\ instance $\mathcal{G}=\tuple{G,\Pi}$. An algorithm for solving the minimization version of \PCEx\ is easily adaptable from Algorithm \ref{alg:pcex}.

\begin{algorithm}
	\caption{\PCEx-BS: a \PCEx\ solver via Binary Search\label{alg:pcex}}
	\textbf{Input:} A coherent \PCE\ instance $\mathcal{G}=\tuple{G,\Pi}$, an action $a_i\in A_i$, and a precision $\varepsilon>0$.
	
	\textbf{Output:} Maximum $P(a_i)$ value with precision $\varepsilon$.
	
	\begin{algorithmic}[1]
		\STATE $k := \lceil|\log\varepsilon|\rceil$;
		\STATE $j := 1$, $v_{min} := 0$, $v_{max} := 1$;
		\IF{$\mathit{PCECoherence}(G,\Pi\cup\{P(a_i)=1\}) = \mathrm{Yes}$}
		\STATE $v_{min} := 1$;
		\ELSE
		\WHILE{$j\leq k$}
		\STATE{$v_{max} = v_{min} + \frac{1}{2^j}$};
		\IF{$\mathit{PCECoherence}(G,\Pi\cup\{P(a_i)\geq v_{max}\}) = \mathrm{Yes}$}
		\STATE $v_{min} := v_{max}$;
		\ENDIF
		\STATE $j\!\!+\!\!+$;
		\ENDWHILE
		\ENDIF
		\RETURN $v_{min}$;
	\end{algorithmic}
\end{algorithm}

For instance, suppose the goal is to find the maximum possible value for constraining $\alpha$: the first iteration consists of solving \PCE\ for $P(\alpha)=1$, if it is coherent, $\overline{P}(\alpha)=1$, if not, $\overline{P}(\alpha)=0$ with precision $2^0$=1. In case the former iteration was not coherent, the second iteration consists of solving \PCE\ for $P(\alpha)=0.5$, if it is coherent, $\overline{P}(\alpha)=0.5$, if not, $\overline{P}(\alpha)=0$, both cases with precision $2^{-1}=0.5$. One more iteration will give precision $2^{-2}=0.25$, and it consists of solving \PCE\ for $P(\alpha)=0.75$ in case the former iteration was coherent, or for $P(\alpha)=0.25$ in case it was not. The process continues until the desired precision is reached and it takes $|\log 2^{-k}|+1= k+1$ iterations to be completed.

\begin{Theorem}\label{thm:pcex}
	Given a precision $\varepsilon>0$, \PCEx\ can be obtained with $O(|\log\varepsilon|)$ iterations of \PCE.
\end{Theorem}

\begin{Example}\label{ex:pcex}
	Suppose we have an observable game $\mathcal{G}_3=\tuple{G_2,\Pi_3}$ with $G_2$ as in Example \ref{ex:gnf} and $\Pi_3$ a set of PCE consisting only of $P(a^2) = 0.9$. In order to solve \PCEx\ for finding $\overline{P}(b^2)$ with precision $2^{-6}$, it will be necessary to solve seven instances of \PCE\ in the form below.
	\begin{align*}
	\pi_2 + \pi_4 + \pi_5 &= 0.9 \\
	\pi_2 + \pi_5 &= p_5 \\
	\pi_1 + \pi_2 + \pi_3 + \pi_4 + \pi_5 &= 1 \\
	\pi_1, \pi_2, \pi_3, \pi_4, \pi_5 &\geq 0
	\end{align*}
	The necessary iterations of \PCE\ are displayed in Table~\ref{tab:approx}. Our algorithm returns $\overline{P}(b^2)\approx 0.890625$, which is accurate within precision $2^{-6}=0.015625$, since $\overline{P}(b^2)=0.9$.
	\begin{table}
		\centering
		\begin{tabular}{c|c|c|c}
			Iteration & $p_5$ & $\pi'$ & Coherence \\
			\hline
			1 & $1_2=1$ & - & No \\
			2 & $0.1_2=0.5$ & $[0.1, 0.5, 0, 0.4, 0]$ & Yes \\
			3 & $0.11_2=0.75$ & $[0.1, 0.75, 0, 0.15, 0]$ & Yes \\
			4 & $0.111_2=0.875$ & $[0.1, 0.875, 0, 0.025, 0]$ & Yes \\
			5 & $0.1111_2=0.9375$ & - & No \\
			6 & $0.11101_2=0.90625$ & - & No \\
			7 & $0.111001_2=0.890625$ & $[0.1, 0.890625, 0, 0.009375, 0]$ & Yes \\
		\end{tabular}
		\caption{Iterations for solving \PCEx\ in Example \ref{ex:pcex}.}
		\label{tab:approx}
	\end{table}
\end{Example}

\section{Coherence allowing mixed equilibria}
\label{sec:pcemix}

We proceed on studying \PCE\ with respect to the more general concept of mixed Nash equilibrium.

A \emph{mixed-strategy} for player $i$ is a rational probability distribution $\sigma_i$ over the set $A_i$ of actions for player $i$, and $\Sigma = \Sigma_1 \times \cdots \times \Sigma_n$ is the set of \emph{mixed-strategy profiles}, in which each $\Sigma_i$ is the set of all possible mixed-strategies for player $i$. It is assumed that each player's choice of strategy is independent from all other players' choices, so the \emph{expected utility function} $U_i$ for player $i$ is given by:
\[U_i(\sigma) = \sum_{a \in A} u_i(a) \prod_{j \in P} \sigma_j(a_j),\]
where $\sigma \in \Sigma$. A mixed-strategy profile $e=\tuple{\sigma_1, \ldots, \sigma_i, \ldots, \sigma_n}$ is a \emph{mixed Nash equilibrium} if, for every player $i$, $U_i(e) \geq U_i(\sigma_1, \ldots, \sigma'_i, \ldots, \sigma_n)$ for every $\sigma'_i \in \Sigma$; each $\sigma_i$ in $e$ is called a \emph{best response} for player $i$ given the other players mixed-strategies in $e$. Then, a mixed-strategy profile is a Nash equilibrium if, and only if, it is composed by best responses for all players. A game $G$ always has at least one mixed Nash equilibrium~\cite{nash51}.

Note that an action profile $a$ may be seen as a mixed-strategy profile $\sigma$ by taking each action $a_i \in a$ for player $i$ as the mixed-strategy $\sigma_i \in \sigma$ that assigns $1$ to $a_i$ and $0$ to the other actions in $A_i$. This way, $a$ is a pure Nash equilibrium if, and only if, its associated $\sigma$ is a mixed Nash equilibrium.

The mixed-strategy setting may be better understood if we think of a game situation that repeatedly occurs and, in each instance, the players choose their actions randomly according to their mixed-strategy.

In this context, an observable game is a game that repeatedly occurs and which is known to be at one of its (mixed) equilibria, but the external observer does not know exactly which one. An instance of the game will be played and the observer assigns subjective probabilities to actions being part of the action profile to be reached in that instance. Formally, an \emph{observable game} is a pair $\mathcal{G}=\tuple{G,\Pi}$ as before. We may again interpret these probability assignments as bets placed by the observer to the actions that the players are allowed to choose.

Another way of understanding observable games is by imagining there are many game situations with the same setting, and that repeatedly occurs, and all these game situations are at some mixed Nash equilibrium. With that knowledge, the external observer looks at one game situation that is about to have a new instance played, but he does not know exactly which game situation among the many ones existing this is. Then, the observer does not know which is the mixed Nash equilibrium the game situation he is looking at is in and he assigns subjective probabilities for the players' actions being part of the action profile resulting from the game situation instance.

As before, we suppose that there is a probability distribution over the mixed-strategy equilibria. It is important to note that this probability distribution is independent from probability distribution in a mixed-strategy. The former probability distribution ranges over mixed-strategy equilibria and the latter ranges over actions. If $e_j=\tuple{\sigma_{1j},\ldots,\sigma_{nj}}\in E_G$, $\sigma_{ij}$ designates the $i$-th component of $e_j$.	In this setting, the probability function $P$ over mixed Nash equilibria induces the probability $P(a_i)$ of an action $a_i$ by
\[P(a_i) = \sum_{e_j \in E_G} \sigma_{ij}(a_i) \cdot P(e_j).\]

The definition of \emph{probabilistic constraints on equilibria (PCE)} is analogous to that in Section \ref{sec:prelim}, namely a set of probability assignments on actions. \PCE\ is similarly defined as the problem of, given an observable game $\mathcal{G}=\tuple{G,\Pi}$, deciding if it is \emph{coherent}, i.e. deciding if there exists a probability function over the set of equilibria that satisfies all constraints in $\Pi$.

\begin{Example}\label{ex:mix}
	Recall Example \ref{ex:game} where we had the game between Alice and Bob in which Alice's actions were $a^1$, $a^2$, and $a^3$, and Bob's actions were $b^1$, $b^2$, and $b^3$, such that the joint utilities are in Table \ref{tab:game}. This game has three pure equilibria, now viewed as special cases of mixed equilibria: $e_1=\tuple{\sigma_1^1,\sigma_2^1}$, where $\sigma_1^1(a^1)=1$, $\sigma_1^1(a^2)=\sigma_1^1(a^3)=0$, $\sigma_2^1(b^1)=1$, $\sigma_2^1(b^2)=\sigma_2^1(b^3)=0$; and $e_2=\tuple{\sigma_1^2,\sigma_2^2}$ and $e_3=\tuple{\sigma_1^3,\sigma_2^3}$ that are also based on the ones described in Example \ref{ex:game}. However, there are several other mixed equilibria among which we highlight $e_4=\tuple{\sigma_1^4,\sigma_2^4}$ given by \[\sigma_1^4(a^1)=\frac{2}{3},~~\sigma_1^4(a^2)=\frac{1}{3},~~\sigma_1^4(a^3)=0,~~~~~~\sigma_2^4(b^1)=\frac{4}{5},~~\sigma_2^4(b^2)=\frac{1}{5},~~\sigma_2^4(b^3)=0.\]
	We have established that if only pure equilibria are considered, the constraints $P(a^2)=\frac{1}{3}$ and $P(b^3)=\frac{1}{4}$ are incoherent. However, in a context that considers mixed Nash equilibria, they are coherent, as we detail in Example \ref{ex:pcemix}.
\end{Example}

\subsection{Complexity of \PCE\ over \dGNP}

We also formulate \PCE\ in linear algebraic terms. Let $\Pi = \{ P(\alpha_i) \bowtie_i p_i, 1 \leq i \leq K \}$ be a set of PCE for an observable game with $M$ mixed Nash equilibria. \PCE\ becomes the problem of deciding the existence of a vector $\pi$ satisfying
\begin{eqnarray}
\nonumber
A \pi &\bowtie& p\\
\label{eq:PCErestrictions_mix}
\mbox{$\sum \pi_j$} &=& 1\\
\nonumber
\pi &\geq& 0
\end{eqnarray}
where $A = [a_{ij}]$ is a $K\times M$ matrix whose columns represent the mixed Nash equilibria in the game. In this case $a_{ij} = \sigma_{pj}(\alpha_i)$, where $p\in P$ is such that $\alpha_i\in A_p$, that is $a_{ij}$ is the probability assignment of action $\alpha_i$ in the mixed equilibrium $e_j$ by player $p$. Again, we join the first two conditions in \eqref{eq:PCErestrictions_mix} in matrix $A$. In this setting, the existence of small witnesses for coherence given by Carath\'eodory's Theorem also applies. This leads to a similar complexity result for \PCE\ as we have in Theorem \ref{thm:nppurenp}.

\begin{Theorem}\label{thm:pcemix}
	\PCE\ over a GNP-class is a problem in NP.
\end{Theorem}

\begin{proof}
	This proof is totally analogous to that of Theorem \ref{thm:nppurenp}. Suppose the observable game $\mathcal{G}=\tuple{G,\Pi}$ is coherent and $|\Pi|=K$. Therefore there exists a probability distribution $\pi$ over the set of all possible (mixed) Nash equilibria that satisfy $\Pi$. By the Carath\'eodory's Theorem~\cite{Eck93} there is a probability distribution assigning non-zero probabilities to at most $K+1$ equilibria. These equilibria are polynomially bounded in size since $G$ is in a GNP-class. Therefore, there is a witness $\pi$ whose size is polynomially bounded attesting $\Pi$ is satisfied, so \PCE\ is in NP.
\end{proof}

It is not known if there exists a non-deterministic polynomial algorithm that computes an exact mixed Nash equilibrium for games with at least $3$ players, and such a result would imply theoretical breakthroughs in complexity theory~\cite{etessami10}. On the other hand, for games with $2$ players an algorithm whose complexity lies in NP has been described~\cite{papadimitriou07}. Thus, restricting \GNPk\ to games with two players generates a GNP-class with respect to mixed Nash equilibrium, which we call \dGNP. Further, we actually have to restrict the equilibrium concept to ``small'' representations, as there are infinitely many mixed Nash equilibria for some games --- i.e. the equilibria representation sizes will be polynomially bounded on the game parameters.

Note that both standard normal form and graphical normal form may be used to represent games in \dGNP. Also note that the restriction on ``small'' representations implies that each game has only finitely many mixed equilibria. The following proposition has a reduction from \cite{conitzer08} that we use for establishing the complexity of \PCE\ over \dGNP\ and the inapproximability results later in this section.

\begin{Proposition}[Conitzer \& Sandholm~\cite{conitzer08}]\label{prop:conitzer}
	Let $\phi$ be a CNF Boolean formula with $n$ propositional variables. Then, there exists a $2$-player game $G_\phi$ which may be built in polynomial time, with $f_i\in A_i$, for $i\in\{1,2\}$, such that $\phi$ is satisfiable if, and only if, it has a mixed Nash equilibrium which is a mixed-strategy profile $\sigma_{SAT}=\tuple{\sigma_1,\sigma_2}$, where $\sigma_1$ assigns positive probability $\frac{1}{n}$ to $n$ distinct actions in $A_1\setminus\{f_1\}$. Furthermore, action profile $\sigma_f=\tuple{f_1,f_2}$ is the only other possible mixed Nash equilibrium in~$G_\phi$.
\end{Proposition}

\begin{Theorem}\label{thm:pcemix2}
	\PCE\ over \dGNP\ is NP-complete.
\end{Theorem}

\begin{proof}
	Membership in NP follows from Theorem \ref{thm:pcemix}. SAT is an NP-complete problem; let us reduce this problem to \PCE. Given a CNF Boolean formula $\phi$, let $G_\phi$ be the game in Proposition \ref{prop:conitzer}. We consider the instance $\mathcal{G}=\tuple{G_\phi,\{P(f_1)=0\}}$ of \PCE, which may be computed from $\phi$ in polynomial time. $\mathcal{G}$ is coherent if, and only if, $G_\phi$ has another mixed Nash equilibrium beyond $\sigma_f$, which happens if, and only if, $\phi$ is satisfiable. Thus, \PCE\ over \dGNP\ is NP-hard.
\end{proof}

\begin{Example} \label{ex:pcemix}
	We show the reduction of \PCE\ to the matrix form \eqref{eq:PCErestrictions_mix} for observable game in Example \ref{ex:mix}. We only show the columns of matrix $A$ corresponding to equilibria mentioned in Example \ref{ex:mix}, which already provides a model satisfying the set of PCE in vector $p$ below.
	\begin{align*}
	A\pi =
	\begin{array}{c}
	\phantom{c} \\ a^1 \\ a^2 \\ a^3 \\ b^1 \\ b^2 \\ b^3
	\end{array}
	\left[
	\begin{array}{cccc}
	1 & 1 & 1 & 1\\
	1 & 0 & 0 & \frac{2}{3}\\
	0 & 1 & 0 & \frac{1}{3}\\
	0 & 0 & 1 & 0\\
	1 & 0 & 0 & \frac{4}{5}\\
	0 & 0 & 0 & \frac{1}{5}\\
	0 & 1 & 1 & 0
	\end{array}
	\right] \cdot \left[
	\begin{array}{c}
	\pi_1 \\ \pi_2 \\ \pi_3 \\ \pi_4
	\end{array}
	\right] = \left[
	\begin{array}{c}
	1 \\ p_2 \\ \frac{1}{3} \\ p_4 \\ p_5 \\ p_6 \\ \frac{1}{4}
	\end{array}
	\right] = p
	\end{align*}
	This matrix system is solvable due to, for example, the vector $\pi = [\frac{1}{2}, \frac{1}{4}, 0, \frac{1}{4}]'$, so the \PCE\ instance is coherent.
\end{Example}

The complexity results of Theorems \ref{thm:pcemix} and \ref{thm:pcemix2} rely on non-deterministic algorithm. For a deterministic algorithm, the immediate possibilities are brute force (exponential) search or reduction to SAT (an NP-complete problem with very efficient implementations).  However, it has been shown that, in the case of probabilistic reasoning, both approaches are far inferior to an approach combining linear algebraic and SAT methods~\cite{FDB2011}.  Thus, we provide a potentially more practical algorithm combining linear algebraic methods and game solvers whose equilibria can be efficiently determined.

\subsection{A column generation algorithm for \PCE\ over \dGNP}

An algorithm for solving \PCE\ has to provide a means to find a solution for Equation \ref{eq:PCErestrictions_mix} if one exists and, otherwise, determine that no solution is possible. Without loss of generality we assume that the first line of matrix $A$ corresponds to the condition $\sum \pi_i = 1$ and that the first position of vector $p$ is $1$ and the remaining positions are sorted in a decreasing order. Note that now matrix $A$ has $K+1$ rows.

Let $\overline{A} = [U_{K+1}|A]$ be the matrix $A$ prefixed with the $\{0,1\}$-upper diagonal square matrix $U_{K+1}$ of dimension $K+1$ in which the positions on the diagonal and above it are $1$ and all the other positions are $0$. Note that both $A$ and $\overline{A}$ can have an exponential number of columns $O(2^K)$.

We now provide a method similar to PSAT-solving to deal with mixed equilibria. Note that in the case of pure equilibria the elements $a_{ij}$ could have values only $0$ and $1$, but now $a_{ij} = \sigma_{pj}(\alpha_i) \in [0,1]$. Also note that the columns of $U_{K+1}$ may not all represent equilibria, so $\overline{A}$ has some columns that represent mixed equilibria and some that do not. We construct a vector $c$ of costs having the same size of matrix $\overline{A}$ such that $c_j \in \{0,1\}$, $c_j=0$ if column $\overline{A}^j$ represents a mixed equilibrium for game $G$. Then we generate the following optimization problem associated to \eqref{eq:PCErestrictions_mix}.
\begin{align} \label{eq:PCEmix_prog}
\begin{array}{lll}
\min 				& c' \cdot \pi \\
\mbox{subject to} 	& \overline{A} \cdot \pi=p \\
& \pi\geq 0
\end{array}
\end{align}
We actually solve the problems where all relations in vector $\bowtie$ in \eqref{eq:PCErestrictions_mix} are ``$=$''. This \emph{standard form} is convenient for using the already existing optimization methods; a formulation for the general case similar to \eqref{eq:PCEmix_prog} may be achieved by performing simple usual linear programming tricks~\cite{bertsimas97}.

\begin{Lemma}\label{thm:PCEmix_prog}
	Given a \PCE\ instance $\mathcal{G}=\tuple{G,\Pi}$ and its associated linear algebraic restrictions \eqref{eq:PCErestrictions_mix}, $\mathcal{G}$ is coherent if, and only if, minimization problem \eqref{eq:PCEmix_prog} has a minimum such that $c'\pi=0$.
\end{Lemma}

Condition $c'\pi=0$ means that only the columns of $\overline{A}^j$ corresponding to actual mixed equilibria of the game $G$ can be attributed probability $\pi_j>0$, which immediately leads to solution of \eqref{eq:PCEmix_prog}. Minimization problem \eqref{eq:PCEmix_prog} can be solved by an adaptation of the simplex method with column generation such that the columns of $A$ are generated on the fly. The simplex method is a stepwise method which at each step considers a basis $B$ consisting of $K+1$ columns of matrix $\overline{A}$ and computes its associated cost~\cite{bertsimas97}. The processing proceeds by finding a column of $\overline{A}$ outside $B$, creating a new basis by substituting one of the columns of $B$ by this new column such that the associated cost never increases. To guarantee the cost never increases, the new column $\overline{A}^j$ to be inserted in the basis has to obey a restriction called reduced cost given by $\tilde{c}_j = c_j-c_B B^{-1} A^j \leq 0$, where $c_j$ is the cost of column $A^j$, $B$ is the basis and $c_B$ is the cost associated to the basis. Note that in our case, we are only inserting columns that represent actual mixed equilibria, so that we only insert columns of matrix $A$ and their associated cost $c_j = 0$. Therefore, every new column $A^j$ to be inserted in the basis $B$ has to obey the inequality
\begin{align}\label{eq:PCEmix_cost}
c_B B^{-1} A^j \geq 0.
\end{align}

A column $A^j$ representing a mixed Nash equilibrium may or may not satisfy condition \eqref{eq:PCEmix_cost}. We call a mixed Nash equilibrium that does satisfy \eqref{eq:PCEmix_cost} as \emph{cost reducing mixed equilibrium}. Our strategy for column generation is given by finding cost reducing mixed equilibrium for a given basis.

\begin{Lemma}\label{thm:nash_cost}
	There exists an algorithm that decides the existence of cost reducing mixed equilibrium whose complexity is in NP.
\end{Lemma}

\begin{proof}
	Since we are dealing with games whose mixed Nash equilibria can be computed in NP, we can guess one such equilibrium and in polynomial time both verify it is a Nash equilibrium for the game and that it satisfies Equation~\ref{eq:PCEmix_cost}.
\end{proof}

We can actually build a deterministic algorithm for Lemma \ref{thm:nash_cost} by reducing it to a SAT problem. In fact, computing equilibrium in \dGNP\ can be encoded in a 3-SAT formula $\phi$; the condition \eqref{eq:PCEmix_cost} can also be encoded by a 3-SAT formula $\psi$ in linear time, e.g. by Warners algorithm~\cite{War1998}, such that the SAT problem consisting of deciding $\phi\cup\psi$ is satisfiable if, and only, if there exists a cost reducing mixed equilibrium. Furthermore its valuation provides the desired column $A^j$. This SAT-based algorithm we call the \PCE\ Column Generation Method.

\begin{algorithm}
	\caption{\PCE-CG: a \PCE\ solver via Column Generation\label{alg:PCEviaCG}}
	\textbf{Input:} A \PCE\ instance $\mathcal{G}=\tuple{G,\Pi}$.
	
	\textbf{Output:} No, if $\mathcal{G}$ is not coherent. Or a solution
	$\tuple{B,\pi}$ that minimizes \eqref{eq:PCEmix_prog}.
	
	\begin{algorithmic}[1]
		\STATE $p := [\{1\} \cup \{p_i ~|~ P(\alpha_i)=p_i \in \Pi, 1 \leq i \leq K\}]$ in descending order; \label{line:iniini}
		\STATE $B^{(0)} := U_{K+1};$ \label{lin:ini}
		\STATE $s := 0$, $\pi^{(s)} = (B^{(0)})^{-1} \cdot p$ and $c^{(s)} = [c_1 \cdots c_{K+1}]';$ \label{line:iniend}
		\WHILE{$c^{(s)}{}' \cdot \pi^{(s)} \neq 0$}
		\label{lin:loop}
		\STATE $y^{(s)} = \mathit{GenerateColumn}(B^{(s)},G,c^{(s)});$ \label{lin:cond}
		\IF{$y^{(s)}$ column generation failed}
		\RETURN No;~~ \label{lin:fail}  \COMMENT{\PCE\ instance is
			unsatisfiable}
		\ELSE
		\STATE $B^{(s+1)} = \mathit{merge}(B^{(s)}, y^{(s)});$ \label{lin:merge}
		\STATE $s\!\!+\!\!+$, recompute $\pi^{(s)} := (B^{(s-1)})^{-1} \cdot p$; $c^{(s)}$ the costs of $B^{(s)}$ columns;
		\ENDIF
		\ENDWHILE\label{lin:endloop}
		\RETURN $\tuple{B^{(s)},\pi^{(s)}}$;~~  \COMMENT{\PCE\ instance is satisfiable} \label{lin:end}
	\end{algorithmic}
\end{algorithm}

Algorithm \ref{alg:PCEviaCG} presents the top level \PCE\ decision procedure. Lines \ref{line:iniini}--\ref{line:iniend} present the initialization of the algorithm. We assume the vector $p$ is in descending order. At the initial step we make $B^{(0)} = U_{K+1}$, this forces $\pi^{(0)}_{K+1} = p_{K+1} \geq 0$, $ \pi^{(0)}_{j} = p_{j} -p_{j+1} \geq 0, 1 \leq j \leq K$; and $c^{(0)} = [c_1 \cdots c_{K+1}]'$, where $c_j=0$ if column $j$ in $B^{(0)}$ is a Nash equilibrium; otherwise $c_j=1$. Thus the initial state $s=0$ is a feasible solution.

Algorithm \ref{alg:PCEviaCG} main loop covers lines \ref{lin:loop}--\ref{lin:endloop} which contains the column generation strategy at beginning of the loop (line \ref{lin:cond}). If column generation fails the process ends with failure in line \ref{lin:fail}; the correctness unsatisfiability by failure is guaranteed by Lemma \ref{thm:PCEmix_prog}. Otherwise a column is removed and the generated column is inserted in a process we called \textit{merge} at line \ref{lin:merge}. The loop ends successfully when the objective function (total cost) $c^{(s)}{}' \cdot \pi^{(s)}$ reaches zero and the algorithm outputs a probability distribution $\pi$ and the set of Nash equilibria columns in $B$, at line \ref{lin:end}.

The procedure \textit{merge} is part of the simplex method which guarantees that given a column $y$ and a feasible solution $\tuple{B,\pi}$ there always exists a column $j$ in $B$ such that if  $B[j:=y]$ is obtained from $B$ by replacing column $j$ with $y$, then there is $\tilde{\pi}\geq 0$ such that $\tuple{B[j:=y],\tilde{\pi}}$ is a feasible solution. We have thus proved the following result.

\begin{Theorem}
	Algorithm \ref{alg:PCEviaCG} decides \PCE\ using column generation.
\end{Theorem}

\subsection{An algorithm for \PCEx\ over \dGNP}

We now analyze \PCEx\ in analogy of what has been done in Section \ref{sec:pcepure}. The definition of \PCEx\ is analogous to that of the pure equilibrium case, i.e. given a coherent observable game $\mathcal{G}=\tuple{G,\Pi}$, with $G$ in \dGNP, and an action $a_i \in A_i$, \PCEx\ consists in finding probability functions $\underline{P}$ and $\overline{P}$  that satisfy $\Pi$ such that $\underline{P}(a_i)$ is minimal and $\overline{P}(a_i)$ is maximal. As far as approximation algorithms are concerned for \PCEx, we have analogous results to Theorems \ref{thm:pureapproxmin} and \ref{thm:pureapproxmax}.

\begin{Theorem}\label{thm:pcemix_inapprox}
	Unless $P=NP$, there does not exist a polynomial time algorithm that approximates, to any precision $\varepsilon\in(0,\frac{1}{2})$, the expected value by the minimization version of \PCEx.
\end{Theorem}

\begin{proof}
	SAT is an NP-complete problem; let us reduce this problem to \PCEx. Given a CNF Boolean formula $\phi$, let $G_\phi$ be the game in Proposition \ref{prop:conitzer}. We consider the coherent observable game $\mathcal{G}=\tuple{G_\phi,\{P(a_i)\geq 0\}}$, for some arbitrary $a_i\in A_i$, for $i\in P$, together with action $f_1$ as an instance of \PCEx. The reduction from $\phi$ to $\mathcal{G}$ may be computed in polynomial time. Suppose there exists a polynomial time algorithm that approximates to precision $\varepsilon\in(0,\frac{1}{2})$ the expected value by the minimization version of \PCEx. If $\phi$ is not satisfiable, the only equilibrium in $G_\phi$ is $\sigma_f$, then $\underline{P}(f_1)=1$ and the supposed algorithm should return $m>1-\varepsilon>\frac{1}{2}$. On the other hand, if $\phi$ is satisfiable, $\underline{P}(f_1)=0$ and the supposed algorithm should return $m<0+\varepsilon<\frac{1}{2}$. Therefore, such algorithm decides an NP-complete problem in polynomial time and $P=NP$.
\end{proof}

\begin{Theorem}\label{thm:pcemix_inapprox2}
	Unless $P=NP$, there does not exist a polynomial time algorithm that approximates, to any precision $\varepsilon\in(0,\frac{1}{6})$, the expected value by the maximization version of \PCEx.
\end{Theorem}

\begin{proof}
	SAT is an NP-complete problem; let us reduce this problem to \PCEx. Given a CNF Boolean formula $\phi$, let $G_\phi$ be the game in Proposition \ref{prop:conitzer}. We consider the coherent observable game $\mathcal{G}=\tuple{G_\phi,\{P(a_2)\geq 0\}}$, for some arbitrary $a_2\in A_2$, together with some arbitrary action $a_1\in A_1\setminus\{f_1\}$ as an instance of \PCEx. The reduction from $\phi$ to $\mathcal{G}$ may be computed in polynomial time. Suppose there exists a polynomial time algorithm that approximates to precision $\varepsilon\in(0,\frac{1}{6})$ the expected value by the maximization version of \PCEx. If $\phi$ is not satisfiable, the only equilibrium in $G_\phi$ is $\sigma_f$, then $\overline{P}(a_1)=0$ and the supposed algorithm should return $M<0+\varepsilon<\frac{1}{6}$. On the other hand, if $\phi$ is satisfiable, $\overline{P}(a_1)=\frac{1}{3}$ and the supposed algorithm should return $M>\frac{1}{3}-\varepsilon>\frac{1}{6}$. Therefore, such algorithm decides an NP-complete problem in polynomial time, hence $P=NP$.
\end{proof}

\PCEx\ in the mixed equilibrium setting may also be solved by Algorithm \ref{alg:pcex} with $\mathit{PCECoherence(G,\Pi)}$ now being a process by Algorithm \ref{alg:PCEviaCG}. We have a similar result as in Theorem~\ref{thm:pcex}.

\begin{Theorem}
	Given an instance of \PCEx\ over \dGNP\ and a precision $\varepsilon>0$, \PCEx\ can be obtained with $O(|\log\varepsilon|)$ iterations of \PCE.
\end{Theorem}

\section{Related work}

The framework defined in Section~\ref{sec:model}, where probabilities are assigned to pure Nash equilibria, is very similar to another concept of equilibrium: the \emph{correlated equilibrium}~\cite{aumann74}. A correlated equilibrium in a game $G=\tuple{P,N,A,u}$ is a probability distribution over the set of action profiles $A$ that satisfies a specific equilibrium property. Despite the similarity, these are distinct objects: while our framework models the uncertainty about which pure Nash equilibrium will be reached in a game (by a probability distribution over $E_G\subset A$), the distribution in a correlated equilibrium is the very concept of equilibrium and is defined over all possible action profiles (not necessarily Nash equilibria).

In a deeper comparison, for both computing a correlated equilibrium and deciding on the coherence of an observable game allowing only pure equilibria, it is necessary to guarantee that a probability distribution on action profiles satisfies some linear inequalities that model the equilibrium property, in the case of correlated equilibrium, and that represents the probabilistic constraints, in the case of coherence. However, while the correlated equilibrium inequalities may be directly derived from the given game~\cite{papadimitriou08}, in order to write the coherence inequalities, it is necessary to compute the Nash equilibria of the game, since the distribution in question is over such equilibria. This difference should explain the discrepancy in complexity between the problem of computing a correlated equilibrium, which is polynomial~\cite{papadimitriou08}, and that of computing a distribution over $E_G$ satisfying a set of PCE, which is non-deterministic polynomial; indeed, the proof we provide for the NP-completeness of \PCE\ (concerning only pure equilibria) depends on the NP-completeness of computing Nash equilibria.

Now, turning to \PCE\ allowing mixed Nash equilibria, there are many results stating that deciding on the existence of mixed equilibria in $2$-player games with some property, such as uniqueness, Pareto-optimality, etc, are NP-complete problems~\cite{GZ89,conitzer08}. \PCE\ may be seen as an addition to this list by Theorem~\ref{thm:pcemix2} in Section~\ref{sec:pcemix} if one glimpses it as the problem of deciding whether there are at most $K+1$ mixed equilibria for which there is an associated vector $\pi$ that satisfies conditions in~\eqref{eq:PCErestrictions_mix}.

\section{Conclusions and future work}

By means of observable games, we have modeled the scenario of uncertainty about exactly what equilibrium is to be reached in a situation that will certainly reach one. To avoid the possible hardness of computing and enumerating all equilibria in order to assign probabilities to them, the proposed model instead allows the observer to assign probabilities to the possible actions of the players involved. \PCE\ over GNP-classes in the pure equilibrium case have been shown to be in NP, and to be NP-complete over NP-complete GNP-classes. We have also provided reductions of \PCE\ to PSAT, and of \PCEx\ to \PCE.

Let us resume the analogy between the study of equilibrium concepts and coherent observable games by highlighting that equilibrium computation is part of an ongoing utility debate~\cite{papadimitriou08}. Since by the aspect of the model \emph{per se}, an equilibrium concept explains agents behavior, the actual computation of an equilibrium might be regarded as completely irrelevant. Nevertheless, it might also be argued that it is only reasonable to accept that agents behave according to an equilibrium if it is not too hard to compute such equilibrium. In light of this discussion, we may conclude that the hardness results concerning \PCE\ point to the difficulty of the observer in being coherent; thus, it may explain failures in the management by local producers when competing with oligopolists.

However, PSAT has been shown to have an easy-hard-easy phase-transition, which means that possibly most cases of PSAT-instances resulting from the reduction from \PCE\ over pure equilibria can be solved easily. By this hypotheses, providing that the reduction itself is not too complex (as in the case of classes \GNP), in most cases it is not difficult for an observer to be coherent and then the responsibility for a poor management falls entirely over the poor knowledge on the oligopolistic market by the local producer.

Moreover, we believe that the framework of observable games we set in this work is in the interest of an observer who actually wants to compute the coherence and the extension of his probabilities over actions which are in equilibrium, independently of whether this equilibrium was actually computed or how it was established, e.g. again the local producer observing an oligopolistic market. In this way, beyond the reduction from \PCE\ to PSAT being encouraging due to the phase transition behavior of PSAT, the improvement in the technologies for implementing linear algebraic solvers and SAT, MAXSAT, and SMT solvers points out that these problems can now be dealt by practical applications in most cases.

For the mixed Nash equilibrium, the possibility of having a general \PCE\ solver lies on very improbable breakthrough in algorithms. However in 2-player games we are in the same zone as for general pure equilibrium \PCE\ problem; since it is NP-complete, most instances are expected to be easily solved in practice.

Finally, we highlight that the optimistic perspectives for solving in practice the problems proposed in this work also justify the uncertainty model that was designed to avoid the need to establish a probability distribution over all equilibria in a game. Solving these problems, of course, involves computing some equilibria, however, our conclusions point out that, in most cases, only a few equilibria need to be computed.

For the future, besides tackling some practical problems and implementations, the techniques here presented can be expanded to other forms of equilibrium such as $\varepsilon$-Nash equilibrium \cite{daskalakis09}.

\section*{Funding}

This study was financed in part by the Coordenação de Aperfeiçoamento de Pessoal de Nível Superior - Brasil (CAPES) - Finance Code 001. Eduardo Fermé was partially supported by NOVA LINCS (UIDB/04516/2020) with the financial support of FCT - Fundação para a Ciência e a Tecnologia, through national funds and  FCT MCTES IC\&DT Project - AAC 02/SAICT/2017. Marcelo Finger was partly supported by Fapesp projects 2019/07665-4, 2015/21880-4 and 2014/12236-1 and CNPq grant PQ 303609/2018-4.

\bibliographystyle{chicago}
\bibliography{mfinger,spreto,new}

\end{document}